\documentclass[letterpaper]{article} 
\usepackage{aaai2026}
\nocopyright
\usepackage{times}  
\usepackage{helvet}  
\usepackage{courier}  
\usepackage[hyphens]{url}  
\usepackage{graphicx} 
\usepackage{natbib}  
\usepackage{caption} 
\usepackage{amsmath}
\usepackage{algorithm}
\usepackage{algorithmic}

\usepackage{newfloat}
\usepackage{listings}
\DeclareCaptionStyle{ruled}{labelfont=normalfont,labelsep=colon,strut=off} 
\floatstyle{ruled}
\newfloat{listing}{tb}{lst}{}
\floatname{listing}{Listing}
\title{You Don't Know Until You Click: Automated GUI Testing for Production-Ready Software Evaluation}

\author{
    Yutong Bian\textsuperscript{1},
    Xianhao Lin\textsuperscript{2},
    Yupeng Xie\textsuperscript{3},
    Tianyang Liu\textsuperscript{4},
    Mingchen Zhuge\textsuperscript{5},\\
    Siyuan Lu\textsuperscript{6},
    Haoming Tang\textsuperscript{1},
    Jinlin Wang\textsuperscript{1},
    Jiayi Zhang\textsuperscript{1,3},
    Jiaqi Chen\textsuperscript{7},\\
    Xiangru Tang\textsuperscript{8},
    Yongxin Ni\textsuperscript{9},
    Sirui Hong\textsuperscript{1},
    Chenglin Wu\textsuperscript{1}\thanks{Chenglin Wu (E-mail: alexanderwu@deepwisdom.ai),
  is the corresponding author.}
}
\vspace{.5em} 
\affiliations{
    \textsuperscript{1}DeepWisdom,
    \textsuperscript{2}Fudan University,
    \textsuperscript{3}HKUST(GZ) 
    \textsuperscript{4}UC San Diego 
    \textsuperscript{5}KAUST,\\
    \textsuperscript{6}Westlake University,
    \textsuperscript{7}Stanford University
    \textsuperscript{8}Yale University,
    \textsuperscript{9}NUS

}

\usepackage{bibentry}

\usepackage{cleveref}
\usepackage{xcolor}
\usepackage{colortbl}  
\usepackage{pifont}
\usepackage{amssymb}

\usepackage{array}          
\usepackage{multirow}       
\usepackage{booktabs}       
\usepackage{makecell}       
\usepackage{longtable}     
\usepackage{adjustbox}     
\usepackage{colortbl}
\usepackage{hhline}
\usepackage{nicematrix}
\usepackage{arydshln} 
\usepackage[most]{tcolorbox}
\usepackage{enumitem}

\definecolor{green(pigment)}{rgb}{0.0, 0.65, 0.31}
\definecolor{darksalmon}{rgb}{0.91, 0.59, 0.48}
\definecolor{champagne}{rgb}{0.97, 0.91, 0.81}

\definecolor{green(pigment)}{rgb}{0.0, 0.65, 0.31}
\definecolor{darksalmon}{rgb}{0.91, 0.59, 0.48}
\newcommand{\Checkmark}{\ding{51}} 
\newcommand{\XSolidBrush}{\ding{55}} 

\definecolor{green(pigment)}{rgb}{0.0, 0.65, 0.31}
\definecolor{darksalmon}{rgb}{0.91, 0.59, 0.48}

\usepackage{xspace}
\newcommand{\ours}{\textbf{RealDevWorld}\xspace}
\newcommand{\benchmark}{{\textbf{RealDevBench}}\xspace}
\newcommand{\evalagent}{AppEvalPilot\xspace}

\definecolor{lightgreen}{rgb}{0.56, 0.93, 0.56}

\begin{document}

\maketitle

\begin{abstract}
    Large Language Models (LLMs) and code agents in software development are rapidly evolving from generating isolated code snippets to producing full-fledged software applications with graphical interfaces, interactive logic, and dynamic behaviors. However, current benchmarks fall short in evaluating such production-ready software, as they often rely on static checks or binary pass/fail scripts, failing to capture the interactive behaviors and runtime dynamics that define real-world usability—qualities that only emerge when an application is actively used. This is the blind spot of current evaluation: \textit{you don't know if an app works until you click through it, interact with it, and observe how it responds.} 
To bridge this gap, we introduce \textbf{RealDevWorld}, a novel evaluation framework for automated end-to-end assessment of LLMs' ability to generate production-ready repositories from scratch. It features two key components: (1) \textbf{RealDevBench}, a diverse collection of 194 open-ended software engineering tasks across multiple domains, incorporating multimodal elements to reflect real-world complexity; and (2) \textbf{AppEvalPilot}, a new agent-as-a-judge evaluation system that simulates realistic, GUI-based user interactions to automatically and holistically assess software functional correctness, visual fidelity, and runtime behavior. The framework delivers fine-grained, task-specific diagnostic feedback, supporting nuanced evaluation beyond simple success/failure judgments. 
Empirical results show that RealDevWorld delivers effective, automatic, and human-aligned evaluations, achieving an accuracy of 0.92 and a correlation of 0.85 with expert human assessments, while significantly reducing the relianc on manual review. This enables scalable, human-aligned assessment of production-level software generated by LLMs. 
Our code is available on GitHub\footnote{https://github.com/tanghaom/AppEvalPilot}.
\end{abstract}

\section{1 Introduction}

\begin{figure}[t!]
    \centering
    \vspace{-1em}
    \includegraphics[width=0.95\columnwidth]
    {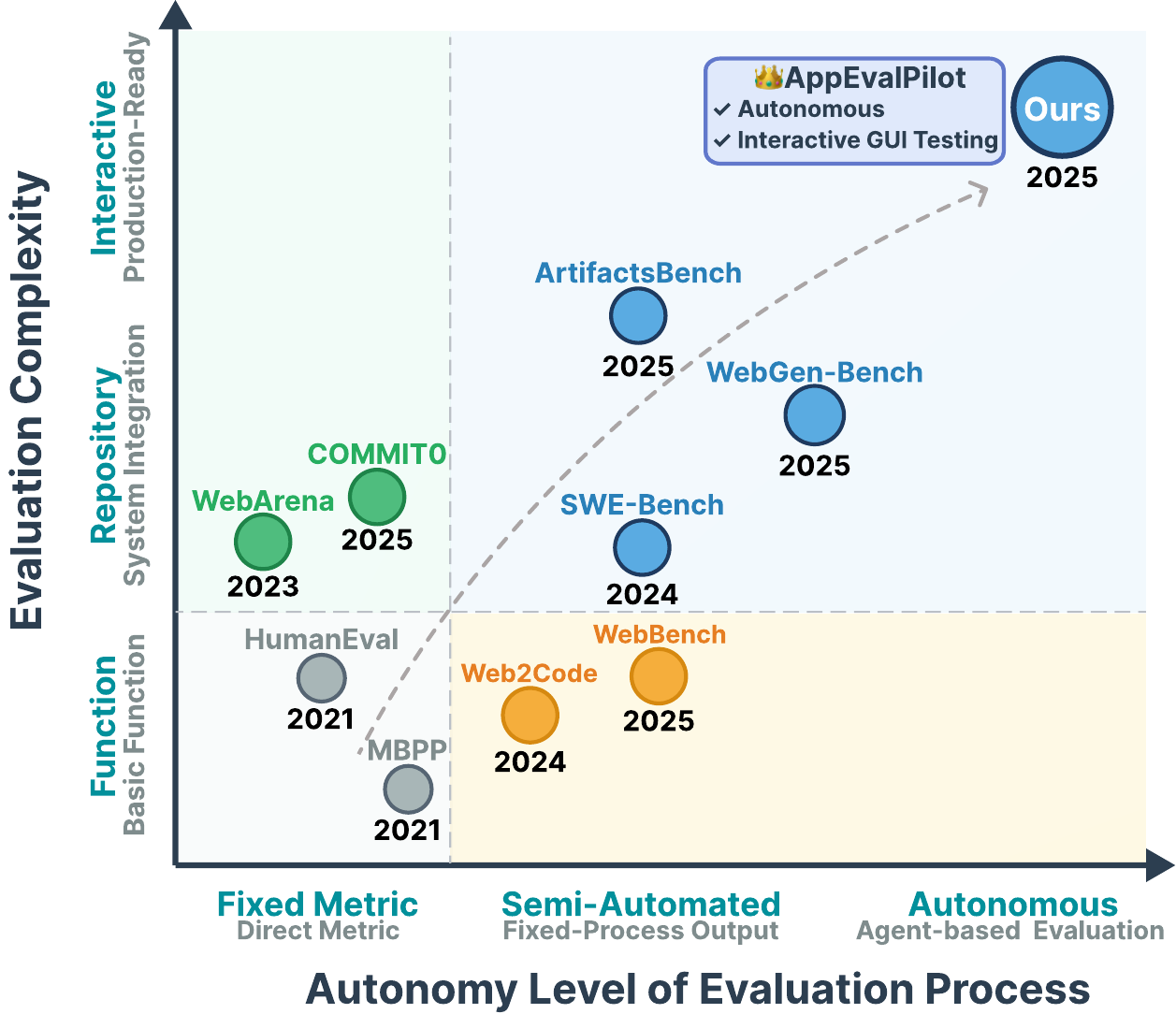}
    \caption{{Software Engineering Evaluation: From Automated to Autonomous Evaluation }}
    \label{fig:software_engineering_evaluation}
    \vspace{-1em}
\end{figure}

Remarkable advancements in LLMs for code and autonomous coding agents are driving a paradigm shift in software development. Their generative capabilities are evolving from function-level code snippets, to crafting self-contained demos, and now towards the creation of sophisticated, production-ready repositories featuring intuitive user interfaces, modular architectures, and robust runtime integration. However, this evolution poses significant challenges for evaluation. Current repository-level code generation tasks lack rigorous assessments of functional completeness, especially with respect to  dynamic and interactive user-centric behaviors. For example, consider a game application generated by such a system. Its correctness and quality cannot be reliably determined by code inspection or static analysis alone. Instead, it requires user-centric validation: clicking through the interface, interacting with game elements, observing state transitions, and receiving feedback in real time—actions that reflect how an actual user would engage with the system. These user-centric and runtime-dependent behaviors are difficult to capture through conventional metrics and often demand the execution of complex end-to-end (E2E) test cases on the generated front-end to assess correctness, interaction quality, and behavioral robustness. However, automating such evaluations remains challenging: generated repositories frequently vary in visual layout, interaction flow, and execution paths, making static or script-based evaluations brittle and often infeasible.

Current benchmarks fall short in automatically assessing the functional completeness and real-world applicability of production-ready repositories, as illustrated in Figure~\ref{fig:software_engineering_evaluation}. Function-level benchmarks~\citep{zhuo2024bigcodebench, jain2025livecodebench, zhang2024naturalcodebench} primarily focus on isolated generation tasks, such as function or class implementation, which fail to capture the complexity and dynamic interactions of real-world repository-level applications. Repository-level benchmarks~\citep{ding2023crosscodeeval, li2025evocodebench, liu2024repobench, zhang2023repocoder, hu2025selfevolving, jimenez2024swebench, miserendino2025swe} attempt to assess entire codebases, yet commonly rely on static or predefined evaluation methods, such as code similarity metrics, unit tests, or scripted integration tests, that are inherently brittle and limited. These methods struggle to reflect real-time interactions, user-driven workflows, runtime errors, or the diverse visual and structural variability of generated outputs. Real-world applications, especially those involving user interfaces, documentation, and multimodal content, exhibit dynamic, unpredictable behaviors. Evaluating them accurately demands intelligent, adaptive methods capable of systematically capturing runtime interaction fidelity and user-centric correctness, highlighting the urgent need for more comprehensive evaluation frameworks.

\begin{table*}[t]
    \centering
    \vspace{-.4cm}

     \renewcommand\tabcolsep{2.1pt}
    \renewcommand\arraystretch{1.1}
    \resizebox{\textwidth}{!}{
        \begin{tabular}{lcccccccc}
            \hline
            \rowcolor{champagne}
            \textbf{Benchmark}                                 & \textbf{Lang.} & \textbf{Level} & \textbf{Tasks} & \textbf{Eval Method} & \textbf{Agent Judge}                   & \textbf{Input Data}                     & \textbf{Interactive}       \\ 
            \hline
            BigCodeBench~\citep{zhuo2024bigcodebench}          & PY             & Func.          & Comp.     & Unit test            & \textcolor{darksalmon}{\XSolidBrush}   & \textcolor{darksalmon}{Text, Code}            & \textcolor{darksalmon}{\XSolidBrush}     \\ 
            \rowcolor{gray!10}
            LiveCodeBench~\citep{jain2025livecodebench}        & PY             & Func.          & Gen.     & Unit test            & \textcolor{darksalmon}{\XSolidBrush}   & \textcolor{darksalmon}{Text, Code}            & \textcolor{darksalmon}{\XSolidBrush}     \\ 
            \rowcolor{gray!10}
            \rowcolor{gray!10}
            RepoBench~\citep{liu2024repobench}                 & PY, Java       & Repo.          & Ret.      & Similarity           & \textcolor{darksalmon}{\XSolidBrush}   & \textcolor{darksalmon}{Text, Code}            & \textcolor{darksalmon}{\XSolidBrush}    \\ 
            SWE-Bench~\citep{jimenez2024swebench}              & PY             & Repo.          & Maint.    & Unit test            & \textcolor{darksalmon}{\XSolidBrush}   & \textcolor{darksalmon}{Multi-modal}      & \textcolor{darksalmon}{\XSolidBrush}     \\ 
            \rowcolor{gray!10}
            EvoCodeBench~\citep{li2025evocodebench}            & PY             & Repo.          & Ret.      & Pass@k               & \textcolor{darksalmon}{\XSolidBrush}   & \textcolor{darksalmon}{Text, Code}            & \textcolor{darksalmon}{\XSolidBrush}    \\ 
            \rowcolor{gray!10}
            SWE-Lancer~\citep{miserendino2025swe}              & JS, TS             & Repo.          & Dev.    & Unit test            & \textcolor{darksalmon}{\XSolidBrush}   & \textcolor{darksalmon}{Multi-modal}            & \textcolor{darksalmon}{\XSolidBrush}    \\ 
            FrontendBench~\citep{zhu2025frontendbench}         & JS              & Repo.          & Gen.    & Unit test            & \textcolor{darksalmon}{\XSolidBrush}   & \textcolor{darksalmon}{Text}            & \textcolor{darksalmon}{\Checkmark}    \\ 
            \rowcolor{gray!10}
            COMMIT0~\citep{zhao2024commit0}                    & PY              & Repo.          & Dev.    & Unit test            & \textcolor{darksalmon}{\XSolidBrush}   & \textcolor{darksalmon}{Multi-modal}            & \textcolor{darksalmon}{\XSolidBrush}    \\ 
            Web-Bench~\citep{xu2025web}                   & JS, TS          & Repo.          & Dev.    & Unit test            & \textcolor{darksalmon}{\XSolidBrush}   & \textcolor{darksalmon}{Text}            & \textcolor{darksalmon}{\XSolidBrush}    \\ 
            \hline
            \ours                                              & PY, JS, TS     & Repo.          & Dev.    & Unit test            & \textcolor{green(pigment)}{\Checkmark} & \textcolor{green(pigment)}{Multi-modal} & \textcolor{green(pigment)}{\Checkmark}   \\ 
            \hline
        \end{tabular}
    }
    \vspace{-.2cm}
    \caption{
        \textbf{Comparison of \ours~with existing benchmarks.}
        It leverages \evalagent for scalable, multi-modal, and interactive software evaluation.
        \textit{Note:  TS = TypeScript; JS = JavaScript; Func. = Function level; Repo. = Repository level; Comp. = Completion; Gen. = Generation; Ret. = Retrieval; Maint. = Maintenance; Dev. = Development.}
    }
    \label{table:baseline_comparison}
\end{table*}

Recent advances in interactive agent technology offer promising directions toward this goal. Emerging paradigms, such as Agent-as-a-Judge~\citep{zhuge2024agent}, employ autonomous agents that execute end-to-end tests by emulating human behaviors, monitoring runtime states, and capturing detailed execution traces. Such agents transcend traditional static metrics, treating evaluated applications not merely as passive test subjects, but as dynamic, interactive environments that inform agent reasoning and decision-making. Building upon this paradigm, we present \textbf{ReaDevWorld}, a comprehensive evaluation framework explicitly designed to assess AI-generated, production-ready codebases through dynamic interaction and open-ended testing scenarios. As part of this framework, we introduce \textbf{RealDevBench}, a benchmark of 194 carefully curated open-ended software engineering tasks across display, analysis, data, and game domains. These tasks are sampled from the real-world programming community requirements and systematically expanded at the function level using LLMs, with a subset incorporating multimodal complexity (structured data, images, audio) to reflect real-world challenges.  To operationalize this benchmark, we develop  \textbf{AppEvalPilot}, a novel agent-based evaluation framework that emulates human interactive software engineering practices. Given a task description and generated code, AppEvalPilot integrates web and OS-level operations to simulate testing workflows, conducting both functional and boundary evaluations for comprehensive software development verification. This agent serves as an automated and effective testbed for production-ready software engineering. 
Our main contributions are:
\begin{itemize}
    \item \textbf{A GUI-Interactive Agent-as-a-Judge Paradigm for Automated Evaluation.} We present AppEvalPilot, a novel agent-as-a-judge evaluation paradigm for production-ready code generation in complex, dynamic interaction scenarios. By simulating realistic user behavior and performing runtime GUI interactions, AppEvalPilot enables fine-grained diagnostics comparable to white-box testing in traditional software engineering.
    \item \textbf{An Open-ended and Scalable Benchmark Suite.} RealDevBench features a diverse set of tasks derived from real-world programming needs, spanning domains like display, analysis, data, and gaming. It benchmarks the ability of code intelligence models to build repository-level software from scratch, with tasks incorporating multimodal inputs—such as images, audio, text, and structured data—to increase reasoning difficulty and scenario realism.
    \item \textbf{Human Alignment and Cost-Effective Validation.} Our framework achieves strong alignment with expert human assessments, reaching an accuracy of $0.92$ and a correlation of $0.85$, substantially outperforming existing automated evaluators. By narrowing the gap between model-based and human evaluation, it enables more reliable and cost-effective validation of generated code.
\end{itemize}
\section{2 Related Work}
\subsection{2.1 Benchmarks for Software Engineering}
Evaluating repository-level code generation in LLM-based agents remains challenging due to the complexity of end-to-end software development, including system integration, dependency management, and dynamic interactions~\citep{zhuge2024agent}. Existing benchmarks such as BigCodeBench~\citep{zhuo2024bigcodebench}, LiveCodeBench~\citep{jain2025livecodebench}, and NaturalCodeBench~\citep{zhang2024naturalcodebench} focus on function- or class-level code completion and rely primarily on static test cases, failing to capture dynamic behaviors like web interfaces or gameplay~\citep{hou2024large, jin2024llms}. As a result, they fall short in assessing real-world development challenges such as integration, ambiguous specifications, and evolving requirements.
Repository-level benchmarks~\citep{ding2023crosscodeeval, li2025evocodebench, liu2024repobench, zhang2023repocoder, hu2025selfevolving, jimenez2024swebench, miserendino2025swe} tackle broader software tasks with interdependent components, but mainly use static metrics like similarity scores or unit tests~\citep{fan2023large, laskar2024systematic}, which may not fully reflect functional correctness. Advanced benchmarks like rSDE-Bench~\citep{hu2025selfevolving}, SWE-Bench~\citep{jimenez2024swebench}, and SWE-Lancer~\citep{miserendino2025swe} depend on pre-defined test cases, limiting their ability to evaluate adaptability to requirement changes or the creation of new modules. DEVAI~\citep{zhuge2024agent} and MLE-Bench~\citep{chan2024mle} introduce automated development tasks for agent evaluation but rely on public datasets, which may be seen during model training.
In contrast, our proposed benchmark supports adaptive module development and dynamic interaction testing, simulating human-like evaluation processes to more comprehensively assess software development capabilities.
\vspace{-.3cm}

\subsection{2.2 Advanced Judgement Approaches}
Recent evaluation techniques have established new paradigms, starting with LLM-as-a-Judge~\citep{zheng2023judging}, which employs language models to evaluate text-based tasks instead of traditional metrics. While effective for textual outputs, this approach is limited to assessing static final result rather than development processes or intermediate outputs.
Agent-as-a-Judge~\citep{zhuge2024agent} builds on this by introducing a dynamic agent-based approach, leveraging multi-dimensional scoring and iterative feedback loops. However, it remains insufficient for evaluating software with complex interactive components, particularly those with GUIs. These require evaluating both interaction flows and the functionality of UI elements, which are more dynamic and nuanced.
To address these challenges, we propose an innovative approach that integrates GUI agent capabilities for interactive testing, inspired by recent advances in GUI agents~\citep{xu2024osagent,cheng2024seeclick}, to mirror human testing processes for a more dynamic and comprehensive evaluation. We summarized the comparisons in Table~\ref{table:baseline_comparison}.

\section{3 Preliminary}\label{sec:preliminary}

{This section formalizes the task of end-to-end software evaluation and analyzes three mainstream evaluation paradigms—human evaluation, LLM-as-a-Judge~\citep{NEURIPS2023_91f18a12}, and Agent-as-a-Judge~\citep{zhuge2024agent}-in terms of their coverage across software quality dimensions, laying the foundation for subsequent experiments and theoretical analysis.}

\subsection{3.1 End-to-End Software Evaluation}

As previously discussed in the introduction, end-to-end testing is essential for assessing production-ready software development. Formally, a generator $\mathcal{A}$ (e.g., a human developer or an AI system) receives a requirement instance $Q=(D, F, M)$, where $D$ is the requirement description, $F$ is the list of desired features, and $M$ represents any supplementary materials. Given this input, the generator is expected to produce a complete software repository $R$.

The goal of end-to-end evaluation is to design an effective method to measure the quality of $R$. Unlike unit testing that focuses on individual components, end-to-end evaluation validates user workflows across all system layers, ensuring the entire software system functions correctly in realistic usage scenarios. This challenge is particularly significant for complex software in real-world, open scenarios, where code structure and interaction are often unpredictable.

\subsection{3.2 Formalization and Evolution of Evaluation Workflows}

According to software engineering standards and validation research (ISO/IEC/IEEE 29119~\cite{iso29119-1:2022}, SV-COMP~\cite{svcomp2024}),
production-grade software must undergo comprehensive validation at three levels: \textbf{unit level} (individual code components), \textbf{system level} (architecture and integration), and \textbf{acceptance level} (user interactions and dynamic behaviors). Only by satisfactorily meeting all three levels can software be deemed production-ready.

We model the end-to-end evaluation process as a unified pipeline that transforms the general evaluation workflow into concrete implementations:
\begin{equation}
(Q, R) \xrightarrow{\text{Identify}}  C \xrightarrow{\text{Execute}} \text{T} \xrightarrow{\text{Judge}} S
\end{equation}
where from task description $Q$ and repository $R$, test cases $C$ are identified, 
These test cases are executed to collect execution traces \textbf{T}, and \textbf{Judge} analyzes these traces to produce the final software quality score $S$. The key differences between evaluation paradigms lie in how test cases $C$ are identified given $Q$ and $R$, how these $C$ are executed to collect traces T, and how Judge analyzes these traces to produce $S$. The three mainstream evaluation paradigms are as follows.

\textbf{Human evaluation workflow:} Human experts participate in the entire process, covering unit, system, and acceptance levels. In this paradigm, experts manually analyze requirement $Q$ and repository $R$, design test cases $C$ based on features $F$. The test cases are executed manually to generate comprehensive $\text{T}_{\text{manual}}$ that covers all validation levels such as unit testing, system testing, and acceptance testing. Subsequently, $\text{Judge}_{\text{human}}$ analyzes the manual traces to produce quality score $S_{\text{manual}}$, e.g. test coverage and pass rates. The advantage is comprehensiveness, but the disadvantage is high cost and low efficiency due to the manual nature of the entire process.

\textbf{LLM-as-a-Judge workflow:} A typical implementation is automatic scoring based on static code analysis (e.g., ArtifactsBench). In this approach, $\text{Execute}_{\text{static}}$ extracts code fragments via fixed scripts or paths, generating limited test cases $C$ only from static code inspection rather than from the original feature list $F$. This produces $\text{Trace}_{\text{static}}$ consisting of static text representations, which $\text{Judge}_{\text{LLM}}$ analyzes through text-based reasoning to generate $Q_{\text{static}}$. This method only covers the unit and part of the system level, cannot detect runtime or interaction issues, and has limited reliability due to the static nature of both $\text{Execute}_{\text{static}}$ and $\text{Trace}_{\text{static}}$.

\textbf{Interactive agent-as-a-judge workflow:} The agent can automatically understand requirements and decompose features from $F$ to generate comprehensive test cases $C$. During evaluation, $\text{Execute}_{\text{agent}}$ executes these $C$ through GUI interactions with $R$, dynamically collecting execution results to form $\text{Trace}_{\text{agent}}$ that captures real-time behaviors and user interactions. $\text{Judge}_{\text{agent}}$ then analyzes these dynamic traces to produce $S_{\text{agent}}$. This method can automatically cover all three dimensions—unit, system, and acceptance levels—combining depth and scalability, making it ideal for production-grade evaluation.
This framework provides the theoretical foundation for our RealDevBench benchmark and AppEvalPilot evaluation system, which we detail in the following sections.

\section{4 \benchmark: Open-Ended SE Benchmark}\label{sec:dataset}

\label{subsec:dataset_overview}

\begin{figure}[t]
    \centering
    \includegraphics[width=\columnwidth]{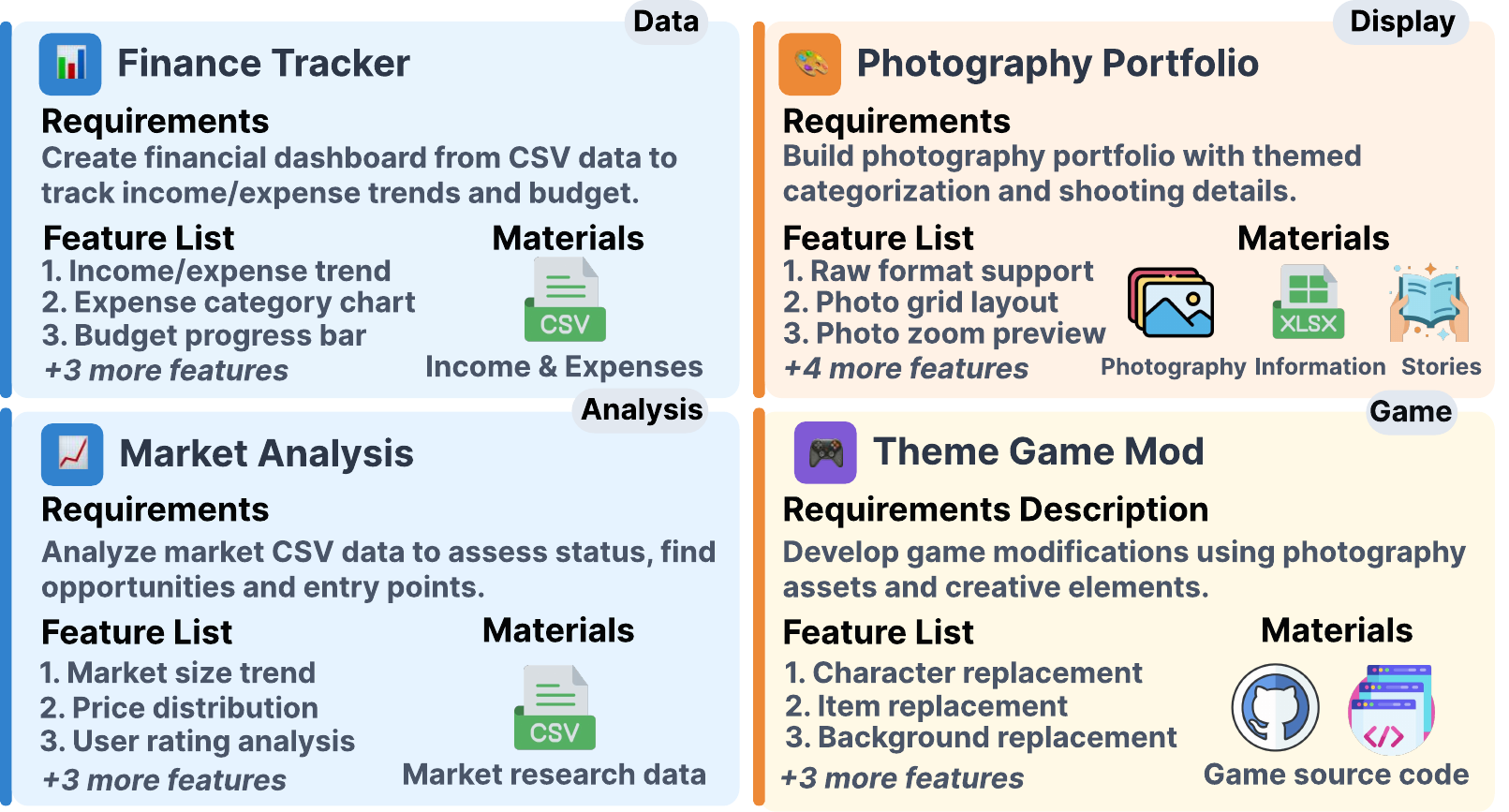}
    \caption{Representative cases from \benchmark across four domains - Data, Display, Analysis, and Game - with consistent triplet structure (requirements, features, materials), reflecting real-world software engineering challenges.}
    \vspace{-1em}
    \label{figure:Cases}
\end{figure}

The \benchmark dataset is constructed through a systematic pipeline designed to ensure its relevance, complexity, and evaluative rigor.

\subsection{4.1 Dataset Overview}
To comprehensively evaluate AI systems across these dimensions, we introduce \benchmark, a benchmark specifically designed to assess end-to-end software engineering capabilities in a realistic and practical context.
\benchmark{} comprises 194 requirements spanning four practical domains—\textit{Analysis, Display, Data}, and \textit{Game}, that reflect core engineering needs. The distribution of tasks is as follows: Display (50.0\%), Data (14.4\%), Analysis (18.6\%), and Game (17.0\%). This allocation mirrors the prevalence of web-centric and data-intensive applications in real-world software development.

The dataset is defined by three key attributes: (1) Open-ended repository construction, where systems must build software from scratch rather than fill in predefined templates; (2) Multimodal complexity, incorporating diverse inputs such as text, images, audio, and tabular data to test integrative and cross-modal capabilities;(3) Functional diversity, encompassing a wide spectrum of software functionalities across varying levels of complexity.

\subsection{4.2 Dataset Construction}

\paragraph{Domain and Requirement.} 
We examined WebDev Arena~\citep{vichare2025webdev} to establish 4 domain categories: \textbf{Display}, \textbf{Analysis} \textbf{Data}, and \textbf{Game}. We sampled requirements from SRDD~\citep{OpenBMB_SRDD} and expanded through web crawling freelancer platforms (Upwork\footnote{https://www.upwork.com} and Freelancer\footnote{https://www.freelancer.com}) to capture real client demands.

\paragraph{Feature Construction.} To construct detailed feature lists that extend requirements from development and functional perspectives, we learned from open-source projects and performed systematic feature extraction. We crawled GitHub projects meeting strict selection criteria: comprehensive documentation (README, API docs), production-ready quality (1000+ stars, active development), and clear feature specifications. We employed Claude-3.5-Sonnet~\citep{anthropic2024claude35sonnet} to extract functional requirements from repository documentation and expand requirements into structured feature specifications, ensuring consistent translation of requirements into actionable development features with clear evaluation criteria.

\paragraph{Task Structure and Formulation.}
\label{subsec:task_structure}
As illustrated in~\Cref{figure:Cases}, each task in \benchmark{} is structured as a triplet to simulate realistic software development scenarios: (1) Requirements Description: A brief textual summary outlining the project's purpose and setting; (2) Feature List: A detailed and structured list of functional goals that define the success criteria; (3) Supplementary Materials: Task-specific resources such as images, audio, or datasets that introduce real-world complexity.

To further enhance the realism of each task, we incorporated carefully curated materials from multiple sources: (1) Images: Sourced from Unsplash~\footnote{https://unsplash.com/} for thematic relevance and professional quality; (2) Datasets: Selected from Kaggle~\footnote{https://www.kaggle.com/} based on topic relevance and appropriate complexity; (3) Documents: Manually created documents (resumes, business proposals, catalogs) that mirror real-world usage scenarios.
\section{5 \evalagent: Autonomous Evaluation}\label{sec:eval_agent}
\begin{figure}
    \centering
    \includegraphics[width=0.5\textwidth, 
    height=6cm, 
    keepaspectratio]{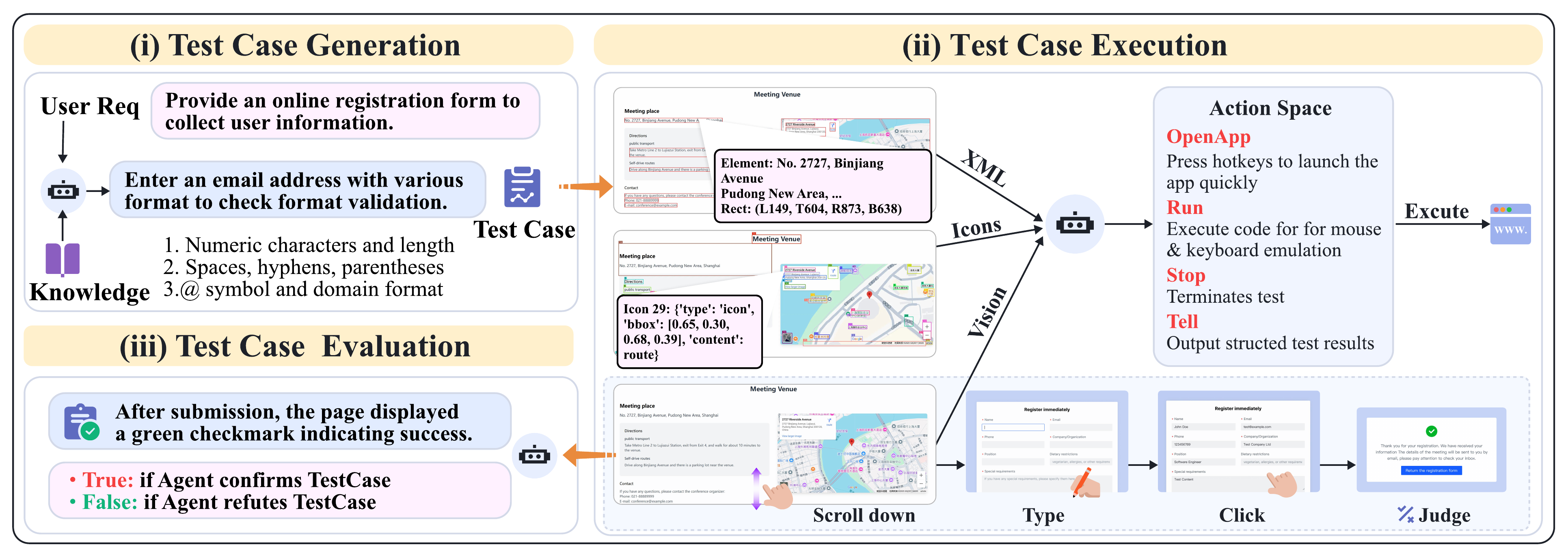}
    \caption{Overall design of \evalagent showing the automated testing workflow: test case generation from user requirements, multimodal test execution through interface interaction (scrolling, typing, clicking), and binary evaluation of outcomes for objective software assessment.}
    \label{EvalAgent overall framework}
    \vspace{-1em}
\end{figure}

As discussed previously, the rise of AI-driven software development demands scalable, automated, and adaptive evaluation methods. To achieve this, we introduce \evalagent, an Agent-as-a-Judge evaluation paradigm designed for automated end-to-end interaction-based software project testing. Unlike static analysis or rigid test suites, \evalagent actively engages with software interfaces, executing real-time user interactions to assess functional correctness and adaptability. As illustrated in Figure~\ref{EvalAgent overall framework}, the evaluation framework follows a three-stage pipeline: (1) generate test cases based on requirements and domain knowledge; (2) simulate real-world user interactions via textual and visual inputs; (3) assess correctness and completeness by comparing actual outcomes with expected behaviors. This dynamic and automated approach aligns with \textbf{RealDevBench}’s focus on practical software evaluation, enabling scalable and rigorous assessment of AI-generated systems.

\paragraph{Test Case Generation.}
\evalagent starts by automating the creation of high-quality, contextually relevant test cases that align with \benchmark’s open-ended and multimodal requirements. 
To achieve this, it leverages few-shot learning ~\citep{wang2020fewshot} to infer requirement-to-test mappings from a small set of manually curated examples, allowing it to generalize efficiently across diverse software requirements.
Additionally, it integrates domain-specific knowledge, such as game mechanics for \textit{Game} tasks, and security protocols for \textit{Data} tasks, to ensure test cases accurately reflect real-world scenarios and practical constraints. 
To standardize generation, the agent uses a structured prompt that simulates the behavior of a professional test engineer. The number of cases is capped (e.g., 15–20) to ensure evaluation tractability.

\begin{figure}
    \centering
    \includegraphics[width=0.5\textwidth, height=6cm, keepaspectratio]{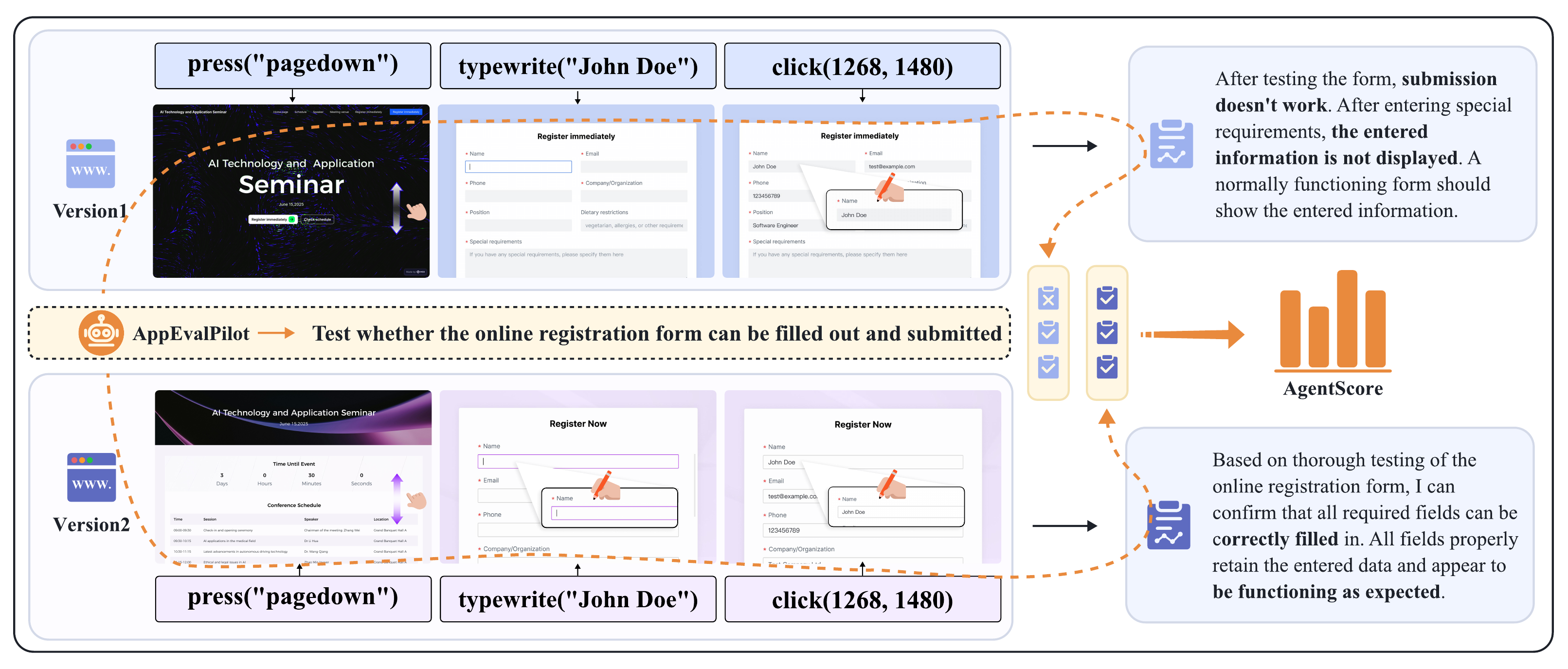}
    \vspace{-.3cm}
    \caption{Evaluation pipeline of \evalagent. The agent performs test sequences on two different web implementations, systematically assesses functionality through direct interaction, documents observable differences in form behavior, and generates quantitative scores based on test cases.}
    
    \label{fig:Evaluate_trajectory}
    \vspace{-1em}
\end{figure}

\paragraph{Test Case Execution.}
\evalagent next autonomously executes the generated test cases by directly interacting with software applications through their graphical user interfaces (GUIs), effectively simulating genuine user interactions.
As shown in \Cref{EvalAgent overall framework}, the execution agent handles multiple input types from active software, including textual data (XML) from accessibility trees (a11ytree) and visual data like icons and screenshots, to accurately interpret the interface.
This facilitates a thorough understanding of the software's UI for precise interaction.
Specifically, the agent operates within a structured action space consisting of four core commands, serving as the foundational components for complex interactions. The action space includes:
\begin{itemize}
    \item Open (app): Launches the target application via shortcut keys to enable quick context switching.
    \item Run (code): Uses PyAutoGUI to simulate mouse and keyboard input for complex interaction sequences.
    \item Tell (answer): Outputs test results to support validation and downstream metrics like \textit{AgentScore}.
    \item \textbf{Stop}: Ends the current test episode, managing execution boundaries.
\end{itemize}
These atomic actions, as shown in~\Cref{EvalAgent overall framework}, allow \evalagent to execute complex tasks such as form filling, web navigation, and validation checks. During the execution of each test case, \evalagent systematically transforms it into a structured, multi-step execution workflow, wherein each step may encompass multiple actions amalgamated to facilitate higher-level operations. To ensure efficiency and flexibility, \evalagent employs adaptive decision-making through historical reasoning and model-based planning, following the Plan-Act framework~\citep{wang2023planandsolve} to continuously improve execution processes. This method allows \evalagent to enhance execution by refining subtasks, minimizing redundant actions, and adapting strategies in response to unexpected UI conditions or errors, especially important for lengthy software testing tasks.

\paragraph{Test Result Evaluation.} \label{subsubsec:test_result_evaluation}
The Test Result Evaluation module compares actual interaction outcomes against the expected success criteria defined in \benchmark. The agent autonomously executes interaction workflows across different application implementations, adapting its actions based on each interface while maintaining consistent testing objectives. Specifically, after each test execution, \evalagent generates a structured report that documents both the performed actions (e.g., entering user information, submitting a form) and the resulting behaviors (e.g., form submission success, data persistence). 
Based on observed outcomes, \evalagent classifies each test case into one of three categories: \textbf{Pass} (expected behavior is met), \textbf{Fail} (expected behavior is violated), or \textbf{Uncertain} (outcome is inconclusive or partially observed). These classifications feed into an aggregated score on test case or feature levels, offering a quantitative assessment of the software quality. 

As illustrated in Figure~\ref{fig:Evaluate_trajectory}, the agent runs similar interaction sequences across different implementations and determines test case satisfaction by comparing observed execution results against specified requirements. This autonomous execution approach enables the agent to make informed judgments about requirement satisfaction by directly observing how different implementations respond to similar user interactions. This process not only surfaces hidden behavioral issues but also ensures that the evaluation remains scalable, interpretable, and grounded in observable user-level feedback.

\section{6 Experiments} 
We conduct comprehensive experiments to validate AppEvalPilot's evaluation capabilities and its effectiveness in benchmarking software development systems. Our experimental design addresses two critical research questions: (1) How effectively does AppEvalPilot evaluate software quality compared to existing evaluation approaches? and (2) Can AppEvalPilot serve as a reliable automated judge for benchmarking LLM-based software engineerinhg?

\subsection{6.1 AppEvalPilot Capability Validation}

\paragraph{Dataset.} We construct our evaluation dataset by selecting 49 tasks (25\%) from RealDevBench, ensuring coverage across all domains. We first fix the generated software projects using Lovable~\citep{lovable} and establish reliable human ground truth labels through a rigorous two-level evaluation process: (1) \textit{Test case-level}: For test cases $c_i$ generated by AppEvalPilot, we invite 3 QA specialists (1-3 years experience) to execute each test case and evaluate Pass/Failed/Uncertain outcomes; (2) \textit{Feature-level}: Each project also receives independent scoring from 3 QA specialists who manually test generated software projects against feature lists, providing granular scores for each feature $f_i \in \{0,1\}$ (Failed/Pass), with final validation by a senior expert. Therefore, each project quality is recorded as $\text{human\_quality} = \frac{1}{n}\sum_{i=1}^{n}f_i$ where $n$ represents the total number of features.

\paragraph{Baselines.} We compare against state-of-the-art GUI systems: Claude-3.5-Sonnet-v2~\citep{anthropic2024claude35sonnet}, UI-Tars~\citep{qin2025uitars}, WebVoyager-Agent~\citep{he2024webvoyager} with qwen2.5-vl-32B~\citep{bai2025qwen2} and claude-3.5-sonnet-v2 backbones, and Browser-Use with claude-3.7-sonnet-v2~\citep{Claude37Sonnet}. Framework protocols provide high-level requirements for autonomous test strategy decomposition, while model protocols provide pre-generated test cases aligned with their input paradigms.

\paragraph{Metrics.} Given test case set $C=\{c_1,c_2,...,c_N\}$ or feature list $F=\{f_1,f_2,...,f_M\}$, each item is classified as true, false, or uncertain by human evaluators or agents. We define binary scores as:
$$\text{score}_{i} = \begin{cases} 
1 & \text{if } \text{class}_{i} = \text{true} \\
0 & \text{if } \text{class}_{i} \in \{\text{false}, \text{uncertain}\}
\end{cases}$$
We use \textbf{accuracy} to measure judgement correctness and \textbf{quality alignment} using Pearson correlation at \textit{test case-level} and \textit{ feature-level}, where test case-level represents averaged performance across all test cases in each project, and feature-level measures correlation between agent and human feature scores across all software projects.
\begin{table}[t]
    \centering

    \renewcommand\tabcolsep{2.6pt}
    \renewcommand\arraystretch{1.2}
    \footnotesize
     \resizebox{0.48\textwidth}{!}{
    \begin{NiceTabular}{@{}l c cc ccc c@{}}[
        code-before = {
            \rowcolor{gray!15}{1,2}  
        }
    ]
        \toprule
        \multirow{2}{*}{\textbf{Method}} 
        & \multicolumn{2}{c}{\textbf{Feature-level}} & \multicolumn{3}{c}{\textbf{Test Case-level}} & \multicolumn{2}{c}{\textbf{Efficiency}} \\
        \cmidrule(lr){2-3} \cmidrule(lr){4-6} \cmidrule(lr){7-8}
        & \textbf{Quality} & \textbf{Align.} & \textbf{Quality} & \textbf{Align.} & \textbf{Acc.}  & \textbf{Time} & \textbf{Cost} \\
        \midrule
        Human  & 0.74 & -- & 0.65 & -- & -- & -- & -- \\
        \midrule
        \multicolumn{8}{c}{\textit{GUI Model}} \\
        \midrule
        Claude-3.5-Sonnet  & 0.27 & 0.23 & 0.46 & 0.49 & 0.68 & 9.20 & 1.01 \\
        UI-Tars  & 0.49 & 0.29 & 0.63 & 0.59 & 0.75 & 8.65 & 0.17 \\
        \midrule
        \multicolumn{8}{c}{\textit{GUI Agent Framework}} \\
        \midrule
        WebVoyager (Qwen2.5)  & 0.29 & 0.25 & 0.35 & 0.44 & 0.6 & 2.16 & \textbf{0.04} \\
        WebVoyager (Claude)  & 0.64 & 0.43 & 0.6 & 0.55 & 0.74 & \textbf{1.60} & 0.10 \\
        Browser-Use (Claude)  & 0.67 & 0.58 & 0.63 & 0.61 & 0.76 & 13.50 & 1.13 \\
        \textbf{AppEvalPilot(Claude)}  & 0.73 & \textbf{0.85} & 0.74 & \textbf{0.81} & \textbf{0.92} & 9.0 & 0.26 \\
        \bottomrule
    \end{NiceTabular}
    }
    \caption{Performance comparison on RealDevBench benchmark. Human Quality (GT) represents ground truth project quality scores from human evaluation. Quality Alignment measures correlation with human assessments.}
    \label{table:realdevbench_results}
\end{table}

\paragraph{Results \& Analysis.} AppEvalPilot demonstrates superior performance across all evaluation metrics. Our framework achieves an accuracy of $0.92$ in test case classification and a quality alignment correlation of $0.81$ with human evaluators, representing a $47\%$ improvement over WebVoyager (Claude-3.5-Sonnet) which achieved $0.55$ accuracy alignment. Compared to baseline GUI testing approaches like Browser-Use~\citep{browser_use2024}, AppEvalPilot reduces evaluation time by $33\%$ (from $13.50$ to $9.00$ minutes per app) while achieving $77\%$ cost reduction through its interactive-driven paradigm. At the feature level, AppEvalPilot maintains the highest alignment with human assessments, achieving $0.85$ correlation across diverse application domains compared to Browser-Use's $0.58$, representing a $47\%$ improvement and validating its effectiveness in end-to-end automated evaluation. End-to-end automated software testing presents significant challenges for existing GUI models and agents, requiring sophisticated planning capabilities and execution accuracy, where traditional GUI tasks primarily focus on fine-grained operational requirements similar to individual test case granularity. When utilizing test cases provided by AppEvalPilot, all baseline models showed an average improvement of $0.17$, demonstrating the value of our test case generation approach. Our observations reveal that detailed test cases not only improve GUI agent testing success rates but also enhance testing robustness, since each feature is decomposed into multiple supporting test cases where incorrect judgment on one test case does not affect the results of other test cases, thereby improving the robustness and reliability of the overall testing process.

\begin{figure}
    \centering
    \includegraphics[width=0.45\textwidth]{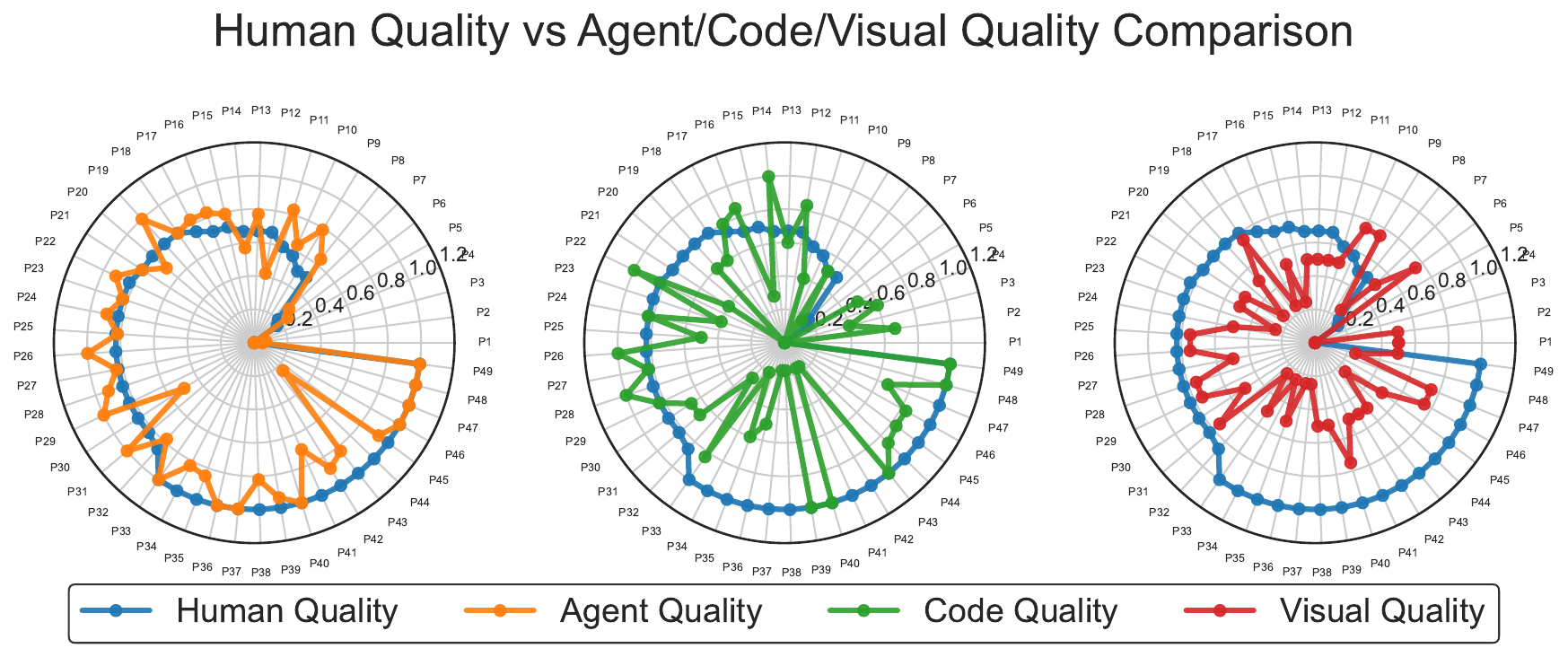}
    \caption{Comparative analysis of evaluation methods versus human quality. (Left) AppEvalPilot's autonomous evaluation, (Middle) Static LLM code scoring, (Right) Visual aesthetic scoring. Each point represents one project, with radial distance indicating quality scores (0-1 scale).}
    \label{fig:radar_comparison}
    \vspace{-1.2em}
\end{figure}

\paragraph{Comparative Evaluation Analysis.} To comprehensively validate AppEvalPilot's evaluation effectiveness, we conduct systematic comparative analysis across multiple evaluation methodologies using the same 49 Lovable-generated projects. Our comparison encompasses static evaluation methods as illustrated in Figure~\ref{fig:radar_comparison}: (1) Code Quality assessment~\citep{NEURIPS2023_91f18a12} employing integrated Claude-3.5-Sonnet scoring of source files, and (2) Visual Quality evaluation utilizing Claude-3.5-Sonnet aesthetic scoring with WebGen-Bench prompts~\citep{lu2025webgenbenchevaluatingllmsgenerating}. As demonstrated in Figure~\ref{fig:radar_comparison}, both Code Quality and Visual Quality fail to effectively capture the nuances of software quality, in contrast to Agent Quality, which shows a strong alignment with human assessments.
Our analysis reveals critical shortcomings in existing LLM-as-a-judge and MLLM-as-a-judge approaches. First, static evaluation cannot capture dynamic interaction issues that define software quality—the deviation means for Code Quality and Visual Quality are 2.79× and 3.34× higher than AppEvalPilot's Agent Quality, respectively, demonstrating substantial gaps between static assessment and actual user experience. Second, evaluation distributions exhibit pronounced misalignment with human judgment: AppEvalPilot achieves a distribution overlap rate of 0.96 with human scores, while Code Quality and Visual Quality achieve mere 0.75 and 0.55 overlap rates, indicating fundamental divergence from natural evaluation patterns.
These findings underscore the superiority of our agent-based evaluation framework in capturing multifaceted software quality aspects that traditional static methods systematically overlook. AppEvalPilot's dynamic interaction capabilities enable accurate quality assessment that closely mirrors human evaluation standards while providing actionable feedback for developers, demonstrating clear advantages over existing static evaluation paradigms.

\subsection{6.2 Performance of LLMs on RealDevBench}
\paragraph{Experimental Setting.} Considering validation costs, we conduct experiments on $54$ tasks from RealDevBench-test. The evaluated generation frameworks include MGX~\citep{mgx}, MGX (BoN-3), Bolt~\citep{bolt}, Lovable, OpenHands~\citep{wang2025openhands}, Claude-3.5-Sonnet, Gemini-2.5-Pro~\citep{comanici2025gemini}, Kimi-K2~\citep{kimiteam2025kimik2openagentic}, DeepSeek-V3~\citep{deepseekai2024deepseekv3technicalreport}, Qwen3-Coder-480B~\citep{qwen3technicalreport}, and Qwen3-235B-Instruct. After code generation, we execute deployment through automated scripts and LLM-generated deployment commands. For MGX, Bolt, and Lovable, we directly utilize their pre-deployed project URLs for testing. We employ three evaluation approaches: AppEvalPilot's interactive assessment, static code quality evaluation, and visual aesthetic scoring through screenshot analysis.

\begin{table}[t]
    \centering

    \renewcommand\arraystretch{1.1}
    \resizebox{0.48\textwidth}{!}{
    \begin{NiceTabular}{l c c c}[
        code-before = {
            \rowcolor{gray!15}{1}  
        }
    ]
        \toprule
        \textbf{System} & \textbf{Agent Quality} & \textbf{Code Quality} & \textbf{Visual Quality} \\
        \midrule
        \multicolumn{4}{c}{\textit{Large Language Models}} \\
        \midrule
        Claude-3.7-Sonnet & 0.31 & 0.41 & 0.18 \\
        Gemini-2.5-Pro & 0.29 & 0.45 & 0.26 \\
        Kimi-K2 & 0.39 & 0.41 & 0.29 \\
        DeepSeek-V3 & 0.29 & 0.18 & 0.21 \\
        Qwen3-Coder-480B & 0.53 & 0.41 & 0.32 \\
        Qwen3-235B-Instruct & 0.33 & 0.42 & 0.20 \\
        \midrule
        \multicolumn{4}{c}{\textit{Agent Systems}} \\
        \midrule
        OpenHands & 0.50 & 0.38 & 0.33 \\
        Lovable & 0.74 & 0.58 & 0.47 \\
        Bolt & 0.54 & 0.69 & \textbf{0.50} \\
        MGX & 0.60 & 0.68 & 0.41 \\
        MGX (BoN-3) & 0.78 & 0.72 & 0.41 \\
        \bottomrule
    \end{NiceTabular}
    }
    
    \caption{Comparative evaluation results across different code generation systems and evaluation methods.}
    \label{table:comparative_evaluation}
    \vspace{-1em}
\end{table}

\paragraph{Performance Analysis.} \benchmark presents significant challenges for LLMs, with even state-of-the-art models like Kimi-K2 achieving only $0.39$ in software quality for generated projects. Current LLM performance on \benchmark is substantially lower than their performance on traditional coding benchmarks, revealing significant defects and bugs in complete interactive functionality development and validation. Visual and static code assessment alone cannot adequately quantify these limitations and shortcomings.
For agent frameworks, generation quality shows significantly higher average scores in Agent Quality, with an improvement of approximately $0.27$ compared to direct LLM generation. This improvement stems from two key factors: First, these frameworks adopt standard software engineering development processes through design, development, and basic deployment verification, significantly enhancing code usability. Second, for complex interactive functionality design, agent-generated projects contain multiple files and components, providing more complete functional implementation compared to single-script solutions produced by LLMs.
As shown in ~\Cref{table:comparative_evaluation},
static assessment methods fail to capture runtime behaviors, user interaction flows, and integration issues that are critical for real-world software functionality. This validates AppEvalPilot's interactive evaluation paradigm as essential for comprehensive software quality assessment.
\section{7 Conclusion}
In this paper, we introduce \textbf{RealDevWorld}, a novel framework for evaluating AI systems that generate code repositories from scratch. It comprises \textbf{RealDevBench}, an open-ended and scalable dataset of 194 diverse tasks with multimodal elements, and \textbf{AppEvalPilot}, a GUI-based Agent-as-a-Judge evaluation paradigm. \textbf{AppEvalPilot} performs automated, end-to-end validation of software functionality, including dynamic behaviors and interaction logic, while providing fine-grained, task-specific diagnostic feedback.

Extensive experiments show that our framework closely aligns with expert human judgments while significantly reducing evaluation time and cost. On the \textbf{RealDevBench} benchmark, \textbf{AppEvalPilot} substantially outperforms existing GUI frameworks, achieving an accuracy of up to 87\%. Overall, \textbf{RealDevWorld} offers a scalable and automated solution for reliable software evaluation, paving the way for future advancements in production-ready code generation.

\bibliography{aaai2026}

\appendix
\clearpage
\onecolumn

\newcommand{\appsection}[1]{%
  \refstepcounter{section}%
  \addcontentsline{toc}{section}{Appendix \Alph{section}\quad #1}%
  \bigskip
  {\Large\bfseries Appendix \Alph{section}\quad #1}\par\nobreak
  \vspace{6pt}  
}

\newcommand{\appsubsection}[1]{%
  \refstepcounter{subsection}%
  \addcontentsline{toc}{subsection}{\Alph{section}.\arabic{subsection}\quad #1}%
  {\large\bfseries \Alph{section}.\arabic{subsection}\quad #1}\par\nobreak
  \vspace{4pt}  
}

\definecolor{myblue}{RGB}{70, 120, 180}
\definecolor{myorange}{RGB}{255,140,0}

\newtcolorbox{displaytaskbox}[2][]{
   enhanced,
  colback=myblue!5,
  colframe=myblue!75!black,
  coltitle=white,
  colbacktitle=myblue!75!black,
  fonttitle=\large\bfseries,
  rounded corners,
  title=#2,
  fontupper=\small,
  toptitle=2pt,
  bottomtitle=2pt,
  top=6pt,
  bottom=6pt,
  left=5pt,
  right=5pt,
  boxsep=3pt,
  drop shadow={black!50!white},
  #1
}

\newtcolorbox{materialsbox}{
  colback=orange!10,
  colframe=orange!60!black,
  rounded corners,
  boxrule=1pt,
  left=8pt,
  right=8pt,
  top=2pt,
  bottom=1pt
}

\appsection{Benchmark Samples Analysis}
\label{appendix:benchmark_samples}

Representative examples from the \benchmark dataset, covering Display, Analysis, Data, and Game domains. These examples illustrate the diverse software engineering challenges evaluated by our benchmark.

\vspace{6pt}
\appsubsection{Display Domain Examples}
The Display domain focuses primarily on web development and content presentation applications requiring front-end development expertise, involving interactive display functionalities such as personal blogs, portfolios, corporate websites, e-commerce storefronts, documentation sites, and multimedia galleries.

\begin{displaytaskbox}{Display Task 1: Professional Portfolio Website}

\textcolor{blue!55!black}{\textbf{\large Requirement Description:}}

\vspace{3pt}

Create a \textbf{professional personal portfolio website} that showcases expertise and project experience. The system will process provided materials to generate a comprehensive, responsive web presence with privacy-conscious content filtering.

\vspace{6pt}

\textcolor{blue!55!black}{\textbf{\large Feature List:}}

\begin{enumerate}[label=\textcolor{blue!55!black}{\textbf{\arabic*.}}, leftmargin=15pt, itemsep=4pt]
  \item \textbf{Navigation System:} Fixed header with smooth scrolling navigation links
  \item \textbf{Hero Section:} Professional profile photograph integration with dynamic introduction
  \item \textbf{Project Showcase:} Interactive card-based layout with hover effects
  \item \textbf{Skills Visualization:} Dynamic skill tag cloud with proficiency indicators
  \item \textbf{Social Integration:} Elegant social media link collection with animations
  \item \textbf{Resume Access:} Secure PDF download with privacy filtering
  \item \textbf{Responsive Design:} Adaptive layout for all device types
\end{enumerate}

\vspace{3pt}

\begin{materialsbox}
\textcolor{orange!75!black}{\textbf{Supplementary Materials:}} 
  Resume document (PDF format), Professional profile photograph (JPG)
\end{materialsbox}

\end{displaytaskbox}

\begin{displaytaskbox}{Display Task 2: Social Link Tree}
\textcolor{blue!55!black}{\textbf{\large Requirement Description:}}
\vspace{3pt}
I have a set of social media links and creative platform homepage links. These materials need to be used to create a link navigation page that conveniently displays all my links on a single page.\\
Please design and implement a social link navigation page based on the following requirements
\vspace{6pt}

\textcolor{blue!55!black}{\textbf{\large Feature List:}}
\begin{enumerate}[label=\textcolor{blue!55!black}{\textbf{\arabic*.}}, leftmargin=15pt, itemsep=4pt]
    \item Display a personal avatar and profile text.
    \item Display all links as a list of buttons.
    \item Links can be filtered by category tags.
    \item Add a theme toggle button to support both light and dark modes.
    \item Generate a QR code for the page to make it easy for others to scan and access.
\end{enumerate}

\vspace{3pt}

\begin{materialsbox}
\textcolor{orange!75!black}{\textbf{Supplementary Materials:}} 
  Link.md containing social media platform URLs
\end{materialsbox}
\end{displaytaskbox}

\appsubsection{Analysis Domain Examples}
The Analysis domain challenges involve transforming raw data into actionable insights:

\begin{displaytaskbox}{Analysis Task 1: Blog Traffic Analysis}
    \textcolor{blue!55!black}{\textbf{\large Requirement Description:}}
\vspace{3pt}
Please design and implement the data analysis based on the following requirements: I have a blog visit data CSV with PV, UV, visit duration, source page, etc. and want to analyze the visit pattern and give optimization suggestions. 
\vspace{6pt}

\textcolor{blue!55!black}{\textbf{\large Feature List:}}
\begin{enumerate}[label=\textcolor{blue!55!black}{\textbf{\arabic*.}}, leftmargin=15pt, itemsep=4pt]
        \item Draw a daily access trend graph to show the trend of blog access.
        \item Provide a ranking of popular articles to show the most visited articles.
        \item Plot the average dwell time graph to analyze how long readers stay on the page.
        \item Provide visit source percentage to help me understand the source channels of visitors.
        \item Provide page bounce rate table to analyze which pages have higher bounce rate.
        \item Provide popular search terms cloud to show the keywords searched by users.    
\end{enumerate}

\vspace{3pt}

\begin{materialsbox}
\textcolor{orange!75!black}{\textbf{Supplementary Materials:}} 
  Blog visit data.csv
\end{materialsbox}
\end{displaytaskbox}

\begin{displaytaskbox}{Analysis Task 2: Product Review Analysis}
    \textcolor{blue!55!black}{\textbf{\large Requirement Description:}}
\vspace{3pt}
Please design and implement the data analysis based on the following requirements. I have a CSV of user review data for a product on an e-commerce platform containing ratings, review text, date of purchase, etc., and would like to analyze these reviews and summarize the product benefits and issues. 

\vspace{6pt}
\textcolor{blue!55!black}{\textbf{\large Feature List:}}
\begin{enumerate}[label=\textcolor{blue!55!black}{\textbf{\arabic*.}}, leftmargin=15pt, itemsep=4pt]
        \item Draw a rating distribution chart to show the distribution of ratings for the product.
        \item Provide a keyword extraction table to analyze the keywords appearing in user reviews.
        \item Plot monthly rating trends and analyze changes in ratings over time.
        \item Provide advantages and problems classification, summarize the advantages and disadvantages of the product.
        \item Provide the rate of favorable and unfavorable charts, showing the proportion of favorable and unfavorable reviews.
        \item Provide an excerpt of popular reviews, showing what users are saying in key reviews.
\end{enumerate}
\vspace{3pt}
\begin{materialsbox}
\textcolor{orange!75!black}{\textbf{Supplementary Materials:}} 
 User comment data.csv
\end{materialsbox}

\end{displaytaskbox}

\newpage
\appsubsection{Data Domain Examples}
The Data domain focuses on information processing and visualization systems:

\begin{displaytaskbox}{Data Task 1: Finance Tracker}
\textcolor{blue!55!black}{\textbf{\large Requirement Description:}}
\vspace{3pt}
Please design and implement the dashboard based on the following requirements: I have a CSV of a year's worth of personal income and expense details, including dates, categories, amounts, notes, and other information. Based on this data, create a personal finance analytics Kanban board that can show income and expenditure trends and track budget execution.

\vspace{6pt}
\textcolor{blue!55!black}{\textbf{\large Feature List:}}
\begin{enumerate}[label=\textcolor{blue!55!black}{\textbf{\arabic*.}}, leftmargin=15pt, itemsep=4pt]
        \item Display a monthly income and expenditure trend chart.
        \item Provide a pie chart of expenditure categories.
        \item Display a budget execution progress bar.
        \item Provide an income and expenditure breakdown grid.
        \item Show a curve of balance changes.
        \item Provides a monthly report out function.
\end{enumerate}
\vspace{3pt}
\begin{materialsbox}
\textcolor{orange!75!black}{\textbf{Supplementary Materials:}} 
Personal income and expenditure details.csv
\end{materialsbox}
\end{displaytaskbox}

\begin{displaytaskbox}{Data Task 2: Stock Data View}
\textcolor{blue!55!black}{\textbf{\large Requirement Description:}}
\vspace{3pt}
Please design and implement the dashboard based on the following requirements: I have a CSV file with historical stock data, including date, opening price, closing price, trading volume, and related news headlines. Based on this data, I would like to create a dashboard to display the market trends of the stock and help me analyze its movement.

\vspace{6pt}
\textcolor{blue!55!black}{\textbf{\large Feature List:}}
\begin{enumerate}[label=\textcolor{blue!55!black}{\textbf{\arabic*.}}, leftmargin=15pt, itemsep=4pt]
        \item Candlestick Chart (K-Line Chart): Display a candlestick chart to visualize the stock's opening, closing, high, and low prices over time.
        \item Trading Volume Bar Chart: Show a bar chart that represents the trading volume on different days.
        \item Technical Indicators Chart: Provide a chart with technical indicators like Moving Averages (MA), Relative Strength Index (RSI), or Bollinger Bands.
        \item News Sentiment Analysis Chart: Display a sentiment analysis chart showing the positive, negative, and neutral sentiment of the related news headlines.
        \item Correlation Heatmap: Provide a heatmap that shows the correlation between the stock price and other related data (such as volume, technical indicators, etc.).
        \item Data Export Feature: Provide a function that allows users to export the analyzed data in a format such as CSV or Excel.
\end{enumerate}
\vspace{3pt}
\begin{materialsbox}
\textcolor{orange!75!black}{\textbf{Supplementary Materials:}} 
Stock historical data.csv
\end{materialsbox}
\end{displaytaskbox}

\appsubsection{Game Domain Examples}
The Game domain challenges test interactive entertainment application development:

\begin{displaytaskbox}{Game Task 1: Mini Card Game}
\textcolor{blue!55!black}{\textbf{\large Requirement Description:}}
\vspace{3pt}
Please develop a card battle game based on the following requirements, where players can play turn-based battles against the computer.

\vspace{6pt}
\textcolor{blue!55!black}{\textbf{\large Feature List:}}
\begin{enumerate}[label=\textcolor{blue!55!black}{\textbf{\arabic*.}}, leftmargin=15pt, itemsep=4pt]
        \item Create a card display interface.
        \item Implement a basic matchmaking system.
        \item Add a simple AI opponent.
        \item Implement a turn counter.
        \item Judge the winners and losers and display the results.
        \item Add a replay button.
\end{enumerate}
\vspace{3pt}
\begin{materialsbox}
\textcolor{orange!75!black}{\textbf{Supplementary Materials:}} 
None
\end{materialsbox}
\end{displaytaskbox}

\begin{displaytaskbox}{Game Task 2: TurboRally Game}
\textcolor{blue!55!black}{\textbf{\large Requirement Description:}}
\vspace{3pt}
Turbo Rally is a racing game software that combines off-road driving with intense rally racing. Players can choose from a variety of rugged vehicles and compete in thrilling rally races on challenging off-road tracks. The objective is to navigate through rough terrain,dodge obstacles,and reach the finish line in the shortest time possible. The game features realistic physics,dynamic weather conditions,and stunning graphics to provide an immersive rally racing experience.  Please design and implement it based on the following requirements:

\vspace{6pt}
\textcolor{blue!55!black}{\textbf{\large Feature List:}}
\begin{enumerate}[label=\textcolor{blue!55!black}{\textbf{\arabic*.}}, leftmargin=15pt, itemsep=4pt]
        \item Implement a vehicle selection interface displaying a minimum of 5 different off-road vehicles with distinct specifications (speed, handling, acceleration) and visual previews
        \item Create a physics engine that simulates realistic vehicle behavior including suspension, terrain interaction, and collision detection with obstacles
        \item Develop a dynamic weather system that affects vehicle handling and track conditions (rain reduces traction, mud affects speed, etc.)
        \item Design a race tracking system that records lap times, checkpoint times, and maintains a leaderboard for each track
        \item Create at least 3 distinct off-road tracks with varying terrain types (mud, gravel, sand) and obstacles (rocks, logs, water crossings)
        \item Implement a real-time performance dashboard showing current speed, lap time, position, and track progress during races
\end{enumerate}
\vspace{3pt}
\begin{materialsbox}
\textcolor{orange!75!black}{\textbf{Supplementary Materials:}} 
None
\end{materialsbox}
\end{displaytaskbox}

\newpage
\definecolor{myblue}{RGB}{70, 120, 180}
\definecolor{myorange}{RGB}{255,140,0}

\newtcolorbox{testcasebox}[2][]{
  enhanced,
  colback=myblue!5,
  colframe=myblue!75!black,
  coltitle=white,
  colbacktitle=myblue!75!black,
  fonttitle=\normalsize\bfseries,
  rounded corners,
  title=#2,
  fontupper=\normalsize,
  toptitle=2pt,
  bottomtitle=1pt,
  top=5pt,
  bottom=5pt,
  left=8pt,
  right=8pt,
  boxsep=1pt,
  drop shadow={black!50!white},
  #1
}

\newtcolorbox{codepromptbox}{
  enhanced,
  colback=gray!5,
  colframe=gray!60!black,
  rounded corners,
  boxrule=1.5pt,
  left=10pt,
  right=10pt,
  top=8pt,
  bottom=8pt,
  boxsep=3pt,
  fontupper=\footnotesize
}

\appsection{AppEvalPilot Details}
\label{appendix:app_eval_pilot}
\vspace{6pt}

\appsubsection{Hierarchical Action Space}
\label{appendix:action_pace}

The action space \(A\) of  \evalagent strikes a balance between expressiveness and operational efficiency, comprising four core actions that enable comprehensive automated testing capabilities, as shown in ~\Cref{table:action_space}. 

\vspace{2pt}

\begin{table}[ht]
    \centering 
    \begin{tabular}{p{0.15\textwidth}p{0.44\textwidth}p{0.3\textwidth}}
        \toprule
        \textbf{Action} & \textbf{Implementation}                                                              & \textbf{Purpose}                      \\ \midrule
        Open (app)      & Using shortcut keys to quickly launch the application (e.g., Win + [Search] + Enter) & Facilitates rapid context switching   \\ \midrule
        Run (code)      & Executes Python scripts via PyAutoGUI for mouse and keyboard emulation               & Enables complex interaction sequences \\ \midrule
        Tell (answer)   & Outputs test results                                                                 & Provides reporting and validation     \\ \midrule
        Stop            & Terminates the test episode                                                          & Controls episode termination          \\ \bottomrule 
    \end{tabular}
     \caption{\textbf{Action Space of \evalagent.} \textit{OpenApp} is designed to facilitate the rapid initialization of the testing environment for \evalagent. The \textit{Run action} constitutes the primary operational module of \evalagent, enabling flexible execution of testing procedures via Python code blocks. The \textit{Tell action} allows \evalagent to output evaluation results. The Stop action terminates the testing process.}
    \label{table:action_space}
\end{table}
\vspace{6pt}

\appsubsection{Agent Execution Case Study}
This section presents case studies designed to demonstrate the agent's ability to evaluate applications across a range of scenarios. Each case study includes the original software design requirements, the corresponding automated test cases, and the agent's evaluation results. For enhanced clarity, screenshots of the agent's operations and historical data are provided at our case study website\footnote{https://appevalpilot.realdev.world}. Analysis of these cases will illustrate \evalagent's dynamic testing capabilities.

\vspace{6pt}
\appsubsection{Prompt for Test Case Generation}
\label{appendix:app_eval_pilot_test_generation}
\begin{testcasebox}{Test Case Examples}
\begin{enumerate}[label=\textcolor{myblue!75!black}{\textbf{\arabic*.}}, leftmargin=18pt, itemsep=3pt]
    \item \textbf{Navigation Verification:} Persistent top navigation bar positioning during scrolling
    \item \textbf{Link Validation:} Intra-page navigation link accuracy ("Home", "Projects", etc.)
    \item \textbf{Image Quality:} Avatar image rendering quality and aspect ratio preservation
    \item \textbf{Content Integrity:} Biographical text completeness and typographic consistency
    \item \textbf{Layout Testing:} Project card list formatting and content integrity
    \item \textbf{Privacy Compliance:} Verify absence of compensation data in project disclosures
    \item \textbf{Responsive Design:} Skill tag cloud layout responsiveness across devices
    \item \textbf{Interactive Elements:} Test hover effects on skill tags and buttons
    \item \textbf{External Links:} Social media link destination accuracy verification
    \item \textbf{Download Function:} PDF resume download functionality testing
    \item \textbf{File Integrity:} Validate PDF file integrity and readability
\end{enumerate}

\end{testcasebox}

\begin{testcasebox}{\textbf{Case Generation Prompt}}
\small
You are a professional test engineer. Please generate a series of specific test cases based on the following user requirements for the webpage.

\vspace{3pt}
\textbf{Requirements:}
\begin{enumerate}[label=\textcolor{myblue!75!black}{\textbf{\arabic*.}}, leftmargin=18pt, itemsep=3pt]
    \item  Test cases must be generated entirely around user requirements, absolutely not missing any user requirements
    \item Please return all test cases in Python list format 
    \item When generating test cases, consider both whether the corresponding module is displayed on the webpage and whether the corresponding function is working properly. You need to generate methods to verify webpage functionality based on your knowledge.
    \item Please do not implement test cases that require other device assistance for verification.
    \item Please control the number of test cases to 15~20, focusing only on the main functionalities mentioned in the user requirements. Do not generate test cases that are not directly related to the user requirements.
    \item When generating test cases, focus on functional testing, not UI testing.
\end{enumerate}

\vspace{3pt}
\texttt{[Test Case Examples]}

\vspace{2pt}
\textbf{User Requirements:} \texttt{[demand]}

\vspace{2pt}
Please return the test case list in List(str) format, without any additional characters, as the result will be converted using the eval function.
\end{testcasebox}
\vspace{6pt}
\appsubsection{Prompt for Test Result Judgment}
\label{appendix:app_eval_pilot_test_judgement}
\vspace{2pt}
\begin{testcasebox}{\textbf{Test Judgement Prompt}}
\small
The model results are labeled as ground truth. Please judge whether the described test case has been successfully implemented based on the facts. If there is evidence that it has been implemented, just output "Yes", otherwise output "No". If the model results indicate that the outcome cannot be determined, output "Uncertain":

\vspace{4pt}
Test Case Description: \texttt{[task\_desc]}

\vspace{4pt}
Model Result: \texttt{[model\_output]}

\vspace{4pt}
Only answer with "Yes", "No", or "Uncertain" 
\end{testcasebox}

\vspace{6pt}

\appsubsection{Prompt for Test Execution}
\label{appendix:app_eval_pilot_test_execution}
\vspace{2pt}
\begin{testcasebox}{\textbf{\large Test Execution Prompt}}
You are a professional and responsible web testing engineer (with real operation capabilities). I will provide you with a test task list, and you need to provide test results for all test tasks. If you fail to complete the test tasks, it may cause significant losses to the client. Please maintain the test tasks and their results in a task list. For test cases of a project, you must conduct thorough testing with at least five steps or more - the more tests, the more reliable the results.

\vspace{4pt}
[IMPORTANT]: You must test ALL test cases before providing your final report! Do not skip any test cases or fabricate results without actual testing! Failing to complete the entire task list will result in invalid test results and significant client losses.

\vspace{4pt}
Task Tips:

Standard Operating Procedure (SOP):

\begin{enumerate}[label=\textcolor{myblue!75!black}{\textbf{\arabic*.}}, leftmargin=18pt, itemsep=3pt]
\item Determine test plan based on tasks and screenshots
\item Execute test plan for each test case systematically - verify each case in the task list one by one
\item After completing each test case, you can use Tell action to report that individual test case result
\item After completing ALL test case evaluations, use Tell action to report the COMPLETE results in the specified format
\end{enumerate}

Reporting Language: Answer in natural English using structured format (like dictionaries). Tell me your judgment basis and results. You need to report the completion status of each condition in the task and your basis for determining whether it's complete.

\vspace{4pt}
Note that you're seeing only part of the app(or webpage) on screen. If you can't find modules mentioned in the task (especially when the right scroll bar shows you're at the top), try using pagedown to view the complete app(or webpage).

\vspace{2pt}
\end{testcasebox}

\vspace{6pt}

\begin{testcasebox}{\textbf{\large Test Execution Report Prompt}}
Inspection Standards:
\begin{enumerate}[label=\textcolor{myblue!75!black}{\textbf{\arabic*.}}, leftmargin=18pt, itemsep=3pt]
\item Test cases are considered Pass if implemented on any page (not necessarily homepage). Please patiently review all pages (including scrolling down, clicking buttons to explore) before ending testing. You must understand relationships between pages - the first page you see is the target app's homepage.
\item If images in tested app(or webpage) modules aren't displaying correctly, that test case fails.
\item You may switch to other pages on the app(or webpage) during testing. On these pages, just confirm the test case result - don't mark other pages-passed cases as Fail if subpages lack features. Return to homepage after judging each case.
\item Trust your operations completely. If expected results don't appear after an operation, that function isn't implemented - report judgment as False.
\item If target module isn't found after complete app(or webpage) browsing, test case result is negative, citing "target module not found on any page" as basis.
\item Don't judge functionality solely by element attributes (clickable etc.) or text ("Filter by category" etc.). You must perform corresponding tests before outputting case results.
\item When tasks require operations for judgment, you must execute those operations. Final results can't have cases with unknown results due to lack of operations (clicks, inputs etc.).
\item For similar test cases (e.g., checking different social media links), if you verify one link works, you can assume others work normally.

\end{enumerate}
\vspace{6pt}
For each individual test case completion, you can use Tell action to report just that result:
\vspace{6pt}
\begin{lstlisting}[
  numbers=none,
  frame=none,
  breaklines=true,
  basicstyle=\scriptsize\ttfamily,
  xleftmargin=0pt
]
Tell ({"case_number": {"result": "Pass/Fail/Uncertain", "evidence": "Your evidence here"}})
\end{lstlisting}

\vspace{6pt}
Even in these failure cases, you must perform sufficient testing steps to prove your judgment before using the Tell action to report all results.

\vspace{6pt}
[VERIFICATION REQUIRED]: Before submitting your final report, verify that:
\begin{enumerate}[label=\textcolor{myblue!75!black}{\textbf{\arabic*.}}, leftmargin=18pt, itemsep=3pt]
\item You have tested EVERY test case in the task list
\item Each test case has an explicit result (Pass/Fail/Uncertain)
\item Each result has supporting evidence based on your actual testing
\end{enumerate}
\vspace{6pt}
Final Result Format (must include ALL test cases):

\begin{lstlisting}[
  numbers=none,
  frame=none,
  breaklines=true,
  basicstyle=\scriptsize\ttfamily,
  xleftmargin=0pt
]
{
    "0": {"result": "Pass", "evidence": "The thumbnail click functionality is working correctly. When clicking on 'Digital Artwork 1' thumbnail, it successfully redirects to a properly formatted detail page containing the artwork's title, image, description, creation process, sharing options, and comments section."},
    "1": {"result": "Uncertain", "evidence": "Cannot verify price calculation accuracy as no pricing information is displayed"},
    "2": {"result": "Fail", "evidence": "After fully browsing and exploring the web page, I did not find the message board appearing on the homepage or any subpage."}
}
\end{lstlisting}

\vspace{6pt}
**Return only the result string. Do not include any additional text, markdown formatting, or code blocks.**

\end{testcasebox}

\vspace{8pt}

\appsubsection{Qualitative Analysis of Failure Modes}
\label{appendix:error_analysis}

To provide a deeper understanding of our agent's behavior, we conducted a qualitative analysis of identified failure cases. This analysis reveals the characteristic limitations of our current approach and provides a roadmap for future improvements in agentic testing. Below, we present a table summarizing common failure modes with specific examples.

{\small
\begin{longtable}{m{0.13\linewidth} m{0.13\linewidth} m{0.13\linewidth} m{0.22\linewidth} m{0.29\linewidth}}

\toprule
\textbf{Project} & \textbf{Test Case} & \textbf{Failure Reason} & \textbf{Analysis} & \textbf{Screenshot} \\
\midrule
\endfirsthead

\multicolumn{5}{c}%
{{\tablename\ \thetable{} -- continued from previous page}} \\
\toprule
\textbf{Project} & \textbf{Test Case} & \textbf{Failure Reason} & \textbf{Analysis} & \textbf{Screenshot} \\
\midrule
\endhead

\endfoot

\bottomrule
\endlastfoot

Language Spelling Bee & Verify that audio playback or definition display for quiz words functions correctly. & Missing Necessary Information & 1.Lack of audio information makes the evidence insufficient. \newline 2. The agent hallucinates a conclusion despite the insufficient evidence. & \includegraphics[width=\linewidth]{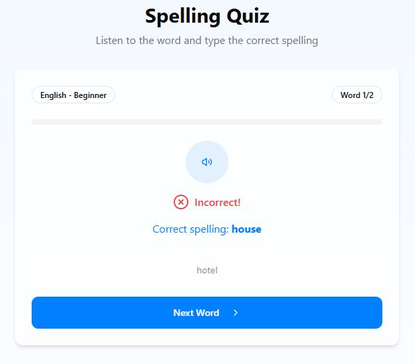} \\
\midrule
MoodMaker & Test if the generated playlist contains between 10-15 songs. & Model Hallucination & The LLM hallucinates. It correctly identifies that there are 3 songs but fails to recognize that 3 is not within the 10-15 range. & \includegraphics[width=\linewidth]{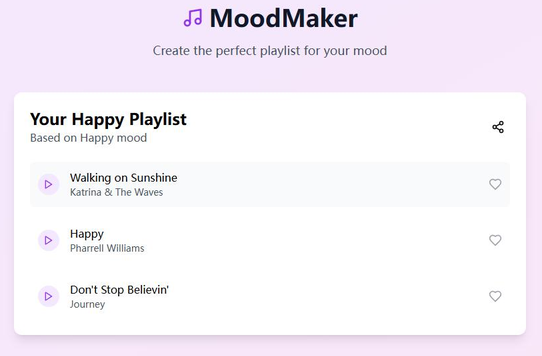} \\
\midrule
Research Paper Gallery & Click on a paper title to check if it navigates to the paper's details page. & Low-quality Test Cases & The generated test case was not aligned with the actual implementation. The "details page" was accessible by clicking "read more," not the title, but the test case was marked as failed for not adhering to the overly specific instruction. & \includegraphics[width=\linewidth]{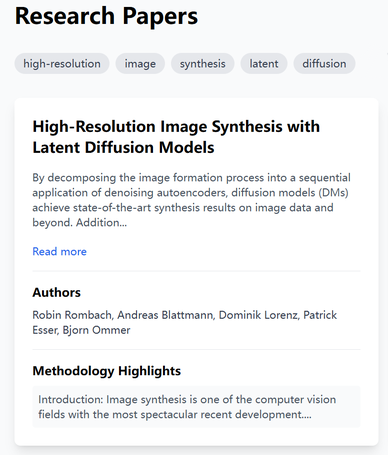} \\
\midrule
Memory Match & Flip all paired cards to verify if the game correctly identifies the completed state. & Need for Advanced Reasoning Ability & The task requires the agent to possess strong logical thinking and memory skills to track and match pairs. & \includegraphics[width=\linewidth]{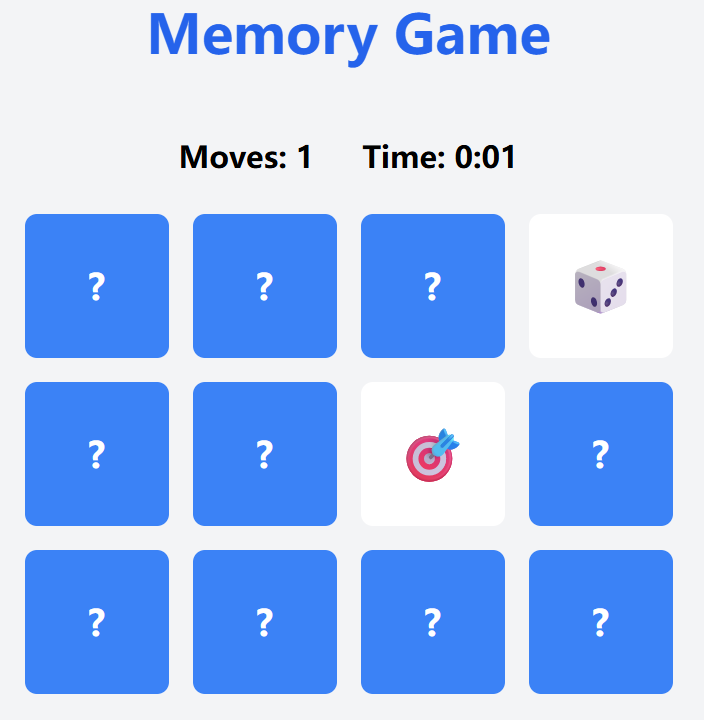} \\
\midrule
Space Shooter & Press the spacebar or the designated shoot key to check if the spaceship fires a projectile. & Need for Real-time Feedback & A significant time lag exists between the agent's observation and its action. By the time the agent decides to act, the environment has already changed. & \includegraphics[width=\linewidth]{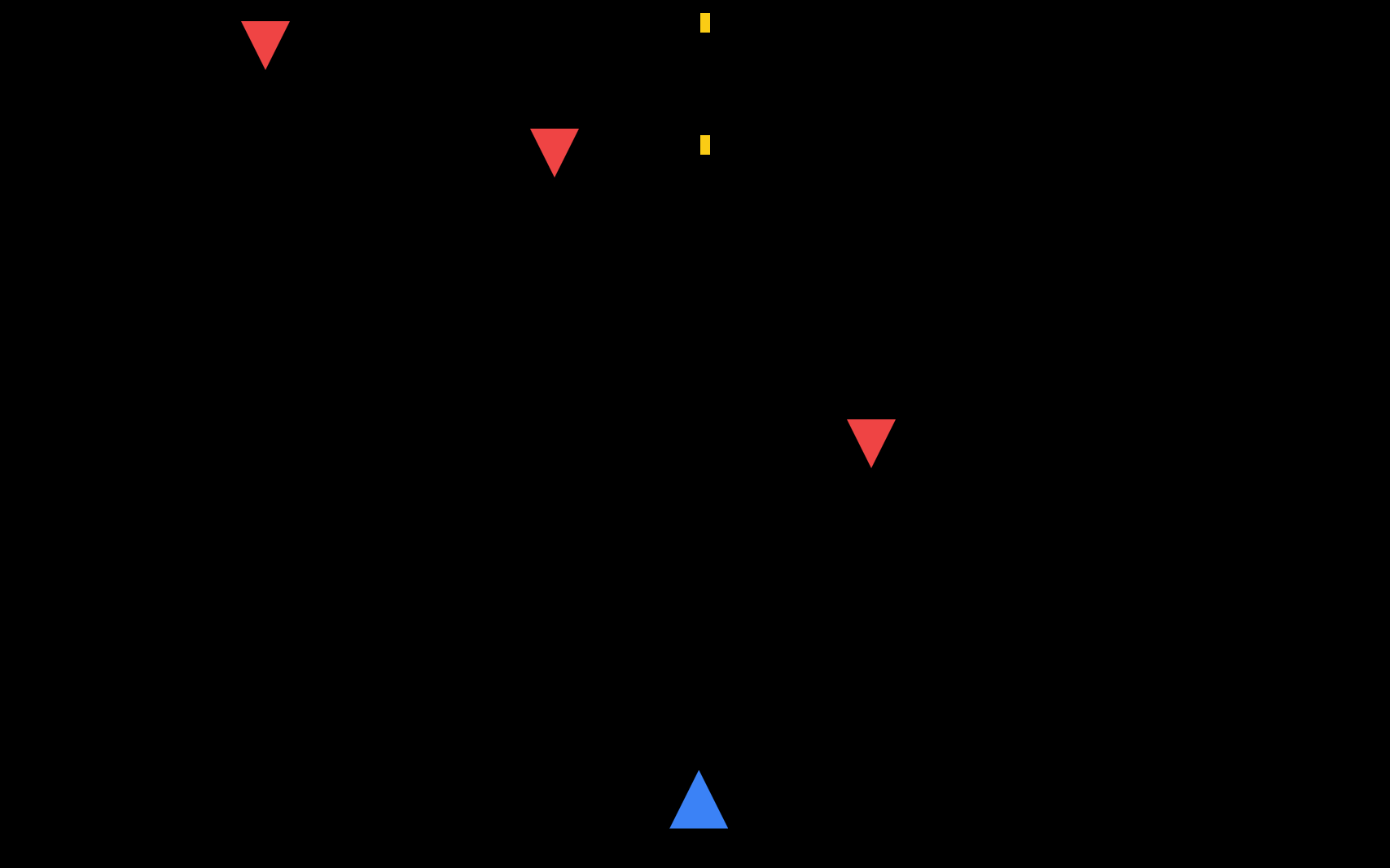} \\
\midrule
Travel Blog & Check if the webpage displays a map module. & Differences in Test Standards Understanding & The agent interpreted the test case literally, passing it as long as a map module was visible. However, the actual requirement implied that the map module must also be fully functional. & \includegraphics[width=\linewidth]{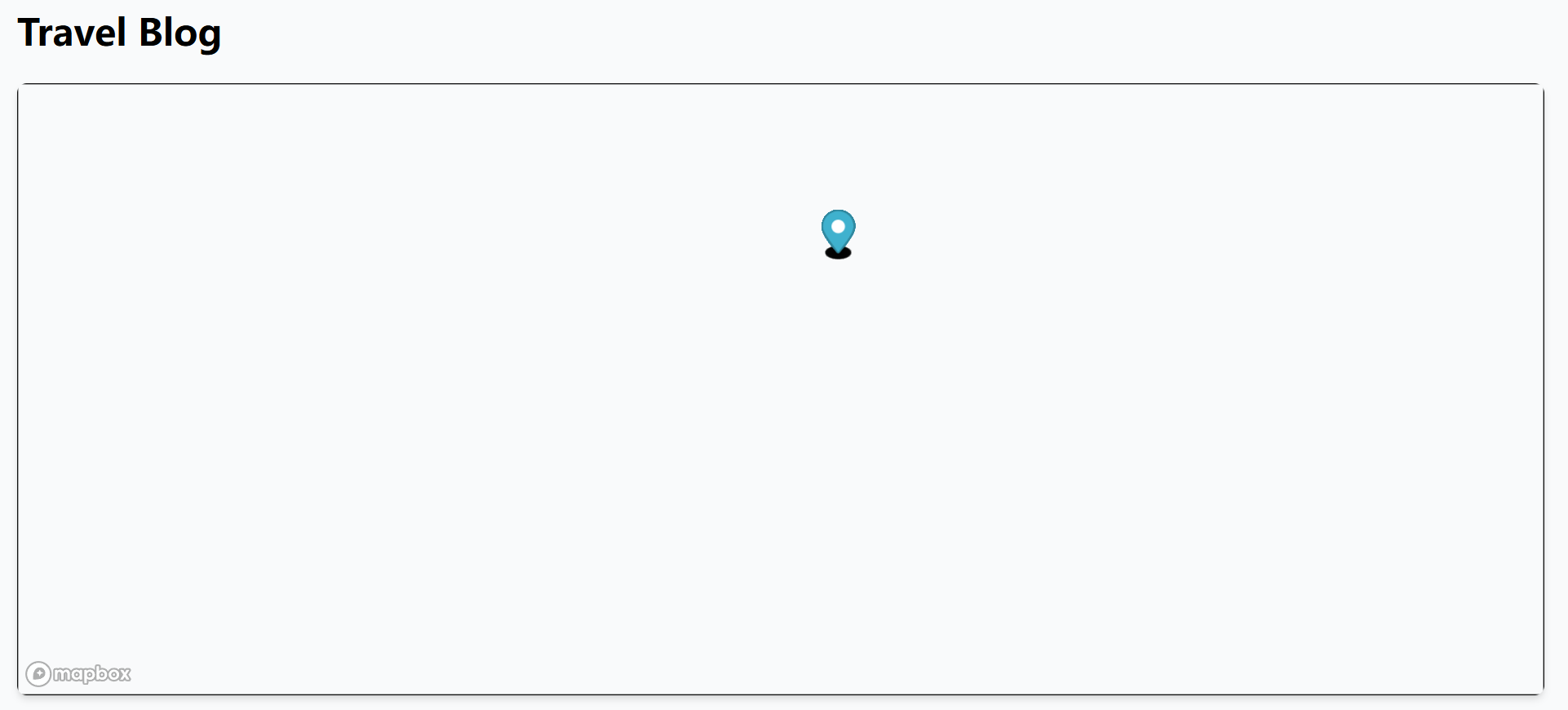} \\

\end{longtable}

\captionof{table}{Examples of Common Failure Modes in Agentic Testing.}
\label{tab:failure_mode_examples_v2}
}

\definecolor{myblue}{RGB}{70, 120, 180}
\newtcolorbox{promptbox}[2][]{
  enhanced,
  colback=myblue!5,
  colframe=myblue!75!black,
  coltitle=white,
  colbacktitle=myblue!75!black,
  fonttitle=\normalsize\bfseries,
  rounded corners,
  title=#2,
  fontupper=\normalsize,
  toptitle=2pt,
  bottomtitle=1pt,
  top=5pt,
  bottom=5pt,
  left=8pt,
  right=8pt,
  boxsep=1pt,
  drop shadow={black!50!white},
  #1
}

\newpage
\appsection{Evaluation of Software Quality}
\label{sq_evaluation}

\appsubsection{Manual Evaluation Process}
\label{manul_sq_evaluation}
In our evaluation process, we invited a total of $12$ individuals, comprising $9$ QA specialists (1-3 years experience) and $3$ senior testing experts (5+ years experience), all of whom are professionals in the field of computer science and experienced software testing engineers. For each project evaluation, we assign $3$ QA specialists to conduct independent assessments, with $1$ senior expert overseeing the quality assurance process. This team conducted comprehensive assessments of each collected task and the generated software projects.

We first fix the generated software projects using Lovable~\citep{lovable} and establish reliable human ground truth labels through a rigorous two-level evaluation process: (1) \textit{Test case-level}: For test cases $c_i$ generated by AppEvalPilot, we invite 3 QA specialists (1-3 years experience) to execute each test case and evaluate Pass/Failed/Uncertain outcomes; (2) \textit{Feature-level}: Each project also receives independent scoring from 3 QA specialists who manually test generated software projects against feature lists, providing granular scores for each feature $f_i \in \{0,1\}$ (Failed/Pass), with final validation by a senior expert. Therefore, each project quality is recorded as $\text{human\_quality} = \frac{1}{n}\sum_{i=1}^{n}f_i$ where $n$ represents the total number of features.

\vspace{4pt}
\textbf{Feature-level Human Annotation Process.} In the feature-level annotation phase, human annotators are required to independently design verification procedures and test cases based on the provided feature list and the implemented application. For each feature $f_i$, annotators create approximately 3-5 test case groups that comprehensively cover different aspects and edge cases of the feature implementation. These test cases are designed to systematically validate whether each feature meets the specified requirements through practical execution scenarios. Based on the execution results of these custom-designed test cases, annotators provide feature implementation labels with three possible outcomes: \textit{true} (feature correctly implemented), \textit{false} (feature failed or incorrectly implemented), and \textit{uncertain} (ambiguous or partially implemented feature requiring further evaluation).

\vspace{4pt}
\textbf{Test Case Annotation Process.} In the test case annotation phase, human annotators are tasked with executing the test cases generated by AppEvalPilot on the implemented applications. For each test case $c_i$, annotators manually perform the specified actions step-by-step on the live application interface. This includes clicking buttons, filling forms, navigating between pages, and triggering interactive elements as described in the test case instructions. During execution, annotators carefully document the complete execution trajectory, recording each action performed, the system's response, and any intermediate states encountered. For each test case, annotators provide comprehensive assessment results including: (1) the complete execution trace documenting each step performed and the corresponding system responses, (2) screenshots or screen recordings of key execution moments, (3) detailed descriptions of any deviations from expected behavior, and (4) a final evaluation label categorized as \textit{true} (test case passed - application behavior matches expectations), \textit{false} (test case failed with clear deviation from expected outcomes), or \textit{uncertain} (ambiguous results requiring expert review, partial functionality or unclear expectations).

\vspace{4pt}
\textbf{Quality Assurance and Validation.} Throughout both annotation phases, all human annotators work independently to ensure unbiased evaluation results. 
To maintain annotation quality and consistency, the assigned senior expert performs comprehensive secondary review of all annotation results for their designated project. This validation process involves identifying cases or features with significant assessment discrepancies across the three independent evaluations, conducting trajectory review and re-execution verification for disputed results, and ensuring the reliability and trustworthiness of the final ground truth labels.

\vspace{4pt}

\appsubsection{Code Quality}
\label{code_quality_evaluation}
\vspace{4pt}
We adopt the Code Quality assessment methodology from~\citep{NEURIPS2023_91f18a12}, employing integrated Claude-3.5-Sonnet for automated scoring of source files. For supported file extensions including \texttt{.py}, \texttt{.html}, \texttt{.css}, \texttt{.js}, \texttt{.ts}, \texttt{.tsx}, and \texttt{.jsx}, we scan the corresponding code files and concatenate their content, then utilize the LLM to evaluate and score the overall code quality. The core prompt is shown as follows.

The evaluation process generates individual scores for each feature in the feature list. We apply a threshold of $75$: features scoring above 75 are marked as passed (1), while those scoring $75$ or below are marked as failed (0). The final LLM score represents the pass rate across all features: $\text{LLM\_score} = \frac{\text{number of passed features}}{\text{total number of features}}$.

\begin{promptbox}{Code Quality Prompt}
\begin{lstlisting}[
  numbers=none,
  frame=none,
  breaklines=true,
  basicstyle=\scriptsize\ttfamily,
  xleftmargin=0pt
]
To perform a comprehensive evaluation of the provided code, focus on a meticulous and step-by-step assessment using the established Software Evaluation Framework, aiming to yield minimal assessment scores based on rigorous real-world high standards.

# Software Evaluation Framework
## Evaluation Criteria
1. Implementation
    - Modularity: Code should be organized into logical, reusable components
    - Architecture: Clear separation of concerns and appropriate design patterns
    - Reusability: Components should be designed for potential reuse
2. Functionality
    - Core Features: All specified features must be fully implemented
    - Interactivity: Dynamic user interactions vs static implementations
    - User Experience: Intuitive and responsive interface
    - Error Handling: Comprehensive error management
    - State Management: Proper handling of application state
3. Logical Flow
    - Control Flow: Clear and efficient program execution paths
    - Data Flow: Proper data transformation and management
    - Event Handling: Appropriate response to system and user events
    - Asynchronous Operations: Proper handling of async processes
    - State Transitions: Clear and predictable state changes
4. Edge Cases
    - Input Validation: Handling of invalid or unexpected inputs
    - Boundary Conditions: Managing edge values and limits
    - Resource Management: Handling resource exhaustion scenarios
5. Requirement Dependencies
    - Feature Dependencies: Proper implementation of dependent features
    - External Services: Correct integration with external services
    - Database Schema: Proper database relationships and constraints

## Quality Metrics and Weightings
### Core Quality Dimensions (Total: 100 points)

1. Functional Correctness (25 points)
2. User Experience (25 points)  
3. Maintainability (20 points)
4. Reliability \& Stability (20 points)
5. Security \& Data Protection (10 points)

# Query
{query}

# Requirements
{features}

# Code
{codes}

# Output Format
Output the evaluation results as a list of Boolean values and corresponding scores in JSON format.
```
[
    {{
        "requirement_id": "Task Id",
        "satisfied": boolean, true or false, satisfies the requirement or not.
        "score": int, 0 ~ 100, the minimal evaluation score based on the high standards.
        "reason": "string, the detailed explanation of the evaluation in 3~5 sentences."
    }},
]
```
## Examples
{example}

# Output
"""
\end{lstlisting}

\end{promptbox}

\end{document}